\documentstyle[12pt]{article}

\begin{document}
\begin{titlepage}
\begin{center}
{\Large\bf On the Role of Higher Twist in \\
\vskip 0.5cm
Polarized Deep Inelastic Scattering}
\end{center}
\vskip 2cm
\begin{center}
{\bf Elliot Leader}\\
{\it Imperial College, Prince Consort Road, London SW7 2BW,
England }
\vskip 0.5cm
{\bf Aleksander V. Sidorov}\\
{\it Bogoliubov Theoretical Laboratory\\
Joint Institute for Nuclear Research, 141980 Dubna, Russia }
\vskip 0.5cm
{\bf Dimiter B. Stamenov \\
{\it Institute for Nuclear Research and Nuclear Energy\\
Bulgarian Academy of Sciences\\
Blvd. Tsarigradsko Chaussee 72, Sofia 1784, Bulgaria }}
\end{center}

\vskip 0.3cm
\begin{abstract}

The higher twist corrections $h^N(x)/Q^2$ to the spin dependent
proton and neutron structure functions $g_1^N(x, Q^2)$ are
extracted in a model independent way from experimental data on
$g_1^N$ and found to be non-negligible. It is shown that the NLO
QCD polarized parton densities determined from the data on $g_1$,
including higher twist effects, are in good agreement with those
found earlier from our analysis of the data on $g_1/F_1$ and $A_1$
where higher twist effects are negligible. On the contrary, the LO
QCD polarized parton densities obtained from the data on $g_1$,
including higher twist, differ significantly from our previous
results.\\

{\it PACS:}13.60.Hb; 13.88+e; 12.38.-t;13.30.-a

\end{abstract}

\end{titlepage}

\newpage
\setcounter{page}{1}

\section{Introduction}

Spurred on by the famous EMC experiment \cite{EMC} at CERN in 1987, there
has been a huge growth of interest in {\it polarized} DIS experiments
which yield more refined information about the partonic structure of the
nucleon, {\it i.e.}, how the nucleon spin is divided up among its
constituents, quarks and gluons. Many experiments have been
carried out at SLAC, CERN and DESY to measure the longitudinal
($A_{\parallel}$) and transverse ($A_{\perp}$) asymmetries and to extract
from them the photon-nucleon asymmetries $A_1(x, Q^2)$ and $A_2(x, Q^2)$
as well as the nucleon spin-dependent structure functions $g_1(x,Q^2)$
and $g_2(x,Q^2)$.
\def\thefootnote{\dagger}
Many theoretical analyses of the world data on $A_1$ and $g_1$
based on leading (LO) and next-to-leading order (NLO)
calculations in perturbative QCD have been performed in order to
test the spin properties of QCD and extract from the data the
polarized parton densities.\footnote{ Note that the theoretical
analyses have been mainly concentrated on the $A_1$($g_1$) data
because the measurements of the quantities $A_2$($g_2$) are much
less accurate with the exception of the very recent data of E155
Collaboration at SLAC \cite{E155A2}. Another reason is that the
theoretical treatment of $g_2$ is much more complicated.} It was
demonstrated that the polarized DIS data are in an excellent
agreement with the pQCD predictions for $A_1^N(x,Q^2)$ and
$g_1^N(x,Q^2)$. What also follows from these analyses is that the
limited kinematic range and the precision of the present
generation of inclusive DIS experiments are enough to determine
with a good accuracy only the polarized parton densities $(\Delta
u +\Delta\bar{u})(x,Q^2)$ and $(\Delta d +\Delta\bar{d})(x,Q^2)$.
The polarized strange sea density $(\Delta s
+\Delta\bar{s})(x,Q^2)$ as well as the polarized gluon density
$\Delta G(x,Q^2)$ are still weakly constrained, especially
$\Delta G$. The non-strange polarized sea-quark densities
$\Delta\bar{u}$ and $\Delta\bar{d}$ cannot be determined, even in
principle, from the inclusive DIS experiments alone without
additional assumptions.

There is, however, an important difference between the kinematic
regions of the unpolarized and polarized data sets. While in the
unpolarized case we can cut the low $Q^2$ and $W^2$ data in order
to eliminate the less known non-perturbative higher twist
effects, it is impossible to perform such a procedure for the
present data on the spin-dependent structure functions without
loosing too much information. This is especially the case for the
HERMES, SLAC and Jefferson Lab experiments. So, to extract the
polarized parton densities from the experimental data on
$g_1^N(x, Q^2)$ the higher twist corrections have to be included
in the data fits. Note that the polarized parton densities in QCD
are related to the leading-twist expression of $g_1$.

It was shown \cite{LomConf,GRSV,LSS2001} that to avoid this
problem and to determine polarized parton densities less
sensitive to higher twist effects it is better to analyze data on
$A_1(\sim g_1/F_1)$  using for the $g_1$ and $F_1$ structure
functions their {\it leading twist} (LT) expressions. It is found
that if for $(g_1)_{\rm LT}$ an NLO approximation is used, the
"effective higher twist" corrections to $A_1$, extracted from the
data, are negligible and consistent with zero within the errors,
which means that the higher twists corrections (HT) to $g_1$ and
$F_1$ approximately cancel in the ratio $g_1/F_1$, or more
precisely, $(g_1)_{\rm HT}/(g_1)_{\rm LT} \approx (F_1)_{\rm
HT}/(F_1)_{\rm LT}$.

In this paper we present a detailed study of the higher twist
contributions $h^N(x)/Q^2$ to the nucleon structure function
$g_1^N(x, Q^2)$. The quantities $h^N(x)$ have been extracted from
the data in a {\it model independent} way. The role of higher
twists in the determination of the polarized parton densities is
discussed.

\section{Connection between Theory and Experiment}

The nucleon spin-dependent structure function $g^N_1(x,Q^2)$ is a
linear combination of the asymmetries $A^N_{\parallel}$ and
$A^N_{\bot}$ (or the related virtual photon-nucleon asymmetries
$A^N_{1}$ and $A^N_{2}$) measured with the target polarized
longitudinally or perpendicular to the lepton beam, respectively.
The most direct way to confront the QCD predictions to the data
is a fit to data on the ratio of the structure functions,
$g_1^N/F_1^N$. Such data have been directly presented by
SLAC/E143 and SLAC/E155 experiments \cite{g1F1data}. Most of the
Collaborations, however, have presented data on the asymmetry
$A_1^N$ which, in practice, are data on $A_{\parallel}^N/D$. The
photon-nucleon asymmetry $A_1^N$ and the ratio $g_1^N/F_1^N$ are
related to the measured longitudinal asymmetry $A^N_{\parallel}$
by
\begin{eqnarray}
A_1^{N}&=&{A_{\parallel}^N\over D}-\eta A_2^N,\\
\label{AparA1A2}
(1+\gamma^2){g_1^{N}\over F_1^{N}}&=&{A_{\parallel}^N\over D}+
(\gamma-\eta)A_2^N,
\label{Aparg1F1}
\end{eqnarray}
where D denotes the photon depolarization factor, $\eta$ and $\gamma$ are
kinematic factors. $\eta$ is proportional to $\gamma$ and $\gamma$ is
given by
\begin{equation}
\gamma^2 = {4M^2_{N}x^2\over Q^2}~.
\label{g2}
\end{equation}
In (\ref{g2}) $\rm M_N$ is the nucleon mass. It should be noted
that in the SLAC and HERMES kinematic regions $\gamma^2$ cannot be
neglected on LHS of (\ref{Aparg1F1}).

The magnitude of $A_2^N$ has been measured by SMC, SLAC/E143 and
SLAC/E155 and found to be small \cite{E155A2,A2data}. Then to a
good approximation its contribution to the RHS of Eqs. (1) and
(\ref{Aparg1F1}) can be neglected and $A_1^N$ and $g_1^N/F_1^N$
can be expressed as
\begin{eqnarray}
A_1^{N}&\cong&{A_{\parallel}^N\over D}~,\\
\label{AparA1appr}
(1+\gamma^2){g_1^{N}\over F_1^{N}}&\cong&{A_{\parallel}^N\over D}~.
\label{Aparg1F1appr}
\end{eqnarray}
It is important to note that due to the additional small factor
$(\gamma-\eta)$ in (\ref{Aparg1F1}) the ratio $g_1^N/F_1^N$ is better
approximated by the measured asymmetry $A_{\parallel}^N$
(Eq. (\ref{Aparg1F1appr})) than the virtual photon-nucleon asymmetry $A_1^N$
(Eq. (4)).

Using (4) and (\ref{Aparg1F1appr}) we reach the well known
relation
\begin{equation}
A_1^{N}(x, Q^2)~ \cong~(1+\gamma^2){g_1^{N}(x, Q^2)\over F_1^{N}(x, Q^2)}
\label{A1g1F1}
\end{equation}
usually used in the literature. However, as was already
mentioned, we have to keep in mind that the presented
experimental values on $A_1^N$ \cite{EMC, A1data} neglecting
$A_2$, $''A_1^N(x, Q^2)_{exp}''$, are really the experimental
values of $A_{\parallel}^N(x, Q^2)/D$ and that the latter quantity
is very well approximated by $(1+\gamma^2)g_1^{N}/ F_1^{N}$.

Using the relation between the unpolarized structure function
$F_1(x,Q^2)$ and the usually extracted from unpolarized DIS
experiments $F_2(x,Q^2)$ and $R(x,Q^2)$
\begin{equation}
2xF^N_1 = F_2^N(1+\gamma^2)/(1 + R^N)~~~~~(N=p,n,d)
\label{ratio}
\end{equation}
Eq. (\ref{A1g1F1}) can be rewritten as
\begin{equation}
A_1^{N}(x,Q^2)\cong {g_1^{N}(x,Q^2)\over
F_2^{N}(x,Q^2)}2x[1+R^{N}(x,Q^2)]~.
\label{assymF2R}
\end{equation}

Up to now, two approaches have been mainly used to extract the
polarized parton densities (PPD) from the world polarized DIS
data. According to the first \cite{GRSV,LSS2001} the leading
twist LO/NLO QCD expressions for the structure functions $g_1^N$
and $F_1^N$ have been used in (\ref{A1g1F1}) in order to confront
the data
\begin{eqnarray}
\nonumber
\left[{g_1(x,Q^2)\over F_1(x, Q^2)}\right]_{exp}~&\Leftrightarrow&~
{g_1(x,Q^2)_{\rm LT}\over F_1(x, Q^2)_{\rm LT}}~,\\
A_1(x,Q^2)_{exp}~&\Leftrightarrow&~(1+\gamma^2){g_1(x,Q^2)_{\rm LT}\over
F_1(x,Q^2)_{\rm LT}}~.
\label{g1F1method}
\end{eqnarray}
In (\ref{g1F1method}) we have dropped the nucleon target label $N$.
It was shown \cite{LomConf,GRSV,LSS2001} that in this case the extracted from
the data ``effective'' HT corrections $h^{A_1}(x)$ to $A_1$
\begin{equation}
A_1(x,Q^2)=(1+\gamma^2){g_1(x,Q^2)_{\rm LT}\over F_1(x,Q^2)_{\rm LT}}
+ {h^{A_1}(x)\over Q^2}
\label{A1HT}
\end{equation}
are negligible and consistent with zero within the errors,
$h^{A_1}(x) \approx 0$ (see Fig. 1). What follows from this
result is that the higher twist corrections to $g_1$ and $F_1$
compensate each other in the ratio $g_1/F_1$ and the PPD
extracted this way are less sensitive to higher twist effects. We
stress again that the polarized parton densities in QCD are
related only to the leading-twist part of $g_1$.

According to the second approach \cite{SMC,BB}, $g_1/F_1$ and $A_1$ data have
been fitted using phenomenological parametrizations of the
experimental data for $F_2(x,Q^2)$ and $R(x,Q^2)$
\begin{eqnarray}
\nonumber
\left[{g_1(x,Q^2)\over F_1(x, Q^2)}\right]_{exp}~&\Leftrightarrow&~
{g_1(x,Q^2)_{\rm LT}\over F_2(x,Q^2)_{exp}}2x{[1+R(x,Q^2)_{exp}]\over
(1+\gamma^2)}~,\\
A_1(x,Q^2)_{exp}~&\Leftrightarrow&~{g_1(x,Q^2)_{\rm LT}\over
F_2(x,Q^2)_{exp}}2x[1+R(x,Q^2)_{exp}]~.
\label{g1F2Rmethod}
\end{eqnarray}
Note that such a procedure is equivalent to a fit to
$(g_1)_{exp}$, but it is in principle better than the fit to the
$g_1$ data themselves actually presented by the experimental
groups. The point is that most of the experimental data on $g_1$
have been extracted from the $A_1$ and $g_1/F_1$ data using the
additional assumption that the ratio $g_1/F_1$ does not depend on
$Q^2$. Also, different experimental groups have used {\it
different} parametrizations for $F_2$ and $R$.

In the analyses \cite{AAC,FS} a procedure has been used which is
somehow a mixture between the two methods above, but bearing in
mind the sensitivity of the results to higher twist effects it is
analogous to the second one. In these fits the leading twist
expression for $F_2$ instead of its experimental values has been
used
\begin{eqnarray}
\nonumber
\left[{g_1(x,Q^2)\over F_1(x, Q^2)}\right]_{exp}~&\Leftrightarrow&~
{g_1(x,Q^2)_{\rm LT}\over F_2(x,Q^2)_{\rm LT}}2x{[1+R(x,Q^2)_{exp}]\over
(1+\gamma^2)}~,\\
A_1(x,Q^2)_{exp}~&\Leftrightarrow&~{g_1(x,Q^2)_{\rm LT}\over
F_2(x,Q^2)_{\rm LT}}2x[1+R(x,Q^2)_{exp}]~.
\label{g1F2Rmix}
\end{eqnarray}

It was shown by GRSV \cite{GRSV} that if the second approach
(\ref{g1F2Rmethod}) is applied to the data ($F_2$ and $R$ are
taken from experiment) allowing at the same time ``effective
higher twist'' contribution $h^{A_1}(x)/Q^2$ to the RHS of
(\ref{g1F2Rmethod}), $h^{A_1}(x)$ is found to be sizeable and
important in the fit. In other words, bearing in mind that a lot
of data on $A_1$ and $g_1/F_1$ are at small $Q^2$ special
attention must be paid to higher twist corrections to the
structure function $g_1$. To extract correctly the polarized
parton densities from the $g_1$ data these corrections have to be
included into data fits. Note that a QCD fit to the data in this
case, keeping in $g_1(x,Q^2)_{QCD}$ only the leading-twist
expression, leads to some "effective" parton densities which
involve in themselves the HT effects and therefore, are not quite
correct.

\section{Higher twist effects in $g_1(x,Q^2)$}

The usual pQCD expression for the nucleon structure function
$g_1^p(x, Q^2)$, in terms of polarized quark and gluon densities,
arises from the contribution of the leading twist ($\tau =2$) QCD
operators and in NLO has the form (a similar formula holds for
$g_1^{n}$):
\begin{equation}
g_1^{p}(x,Q^2)_{\rm pQCD}={1\over 2}\sum _{q} ^{N_f}e_{q}^2
[(\Delta q +\Delta\bar{q})\otimes (1 + {\alpha_s(Q^2)\over
2\pi}\delta C_q) +{\alpha_s(Q^2)\over 2\pi}\Delta G\otimes
{\delta C_G\over N_f}], \label{g1partons}
\end{equation}
where $\Delta q(x,Q^2), \Delta\bar{q}(x,Q^2)$ and $\Delta
G(x,Q^2)$ are quark, anti-quark and gluon polarized densities in
the proton, which evolve in $Q^2$ according to the spin-dependent
NLO DGLAP equations. $\delta C(x)_{q,G}$ are the NLO
spin-dependent Wilson coefficient functions and the symbol
$\otimes$ denotes the usual convolution in Bjorken $x$ space.
$\rm N_f$ is the number of active flavors. In LO QCD the
coefficients $\delta C(x)_{q}$ and $\delta C(x)_{G}$ vanish and
the polarized parton densities in (\ref{g1partons}) evolve in
$Q^2$ according to the spin-dependent LO DGLAP equations.

\def\thefootnote{\dagger}
It is well known that at NLO and beyond, the parton densities as
well as the Wilson coefficient functions become dependent on the
renormalization (or factorization) scheme employed.\footnote{Of
course, physical quantities such as the virtual photon-nucleon
asymmetry $A_1(x,Q^2)$ and the polarized structure function
$g_1(x,Q^2)$ are independent of choice of the factorization
convention.} Two often used schemes are the $\rm \overline{MS}$
and the JET schemes. Both the NLO polarized coefficient functions
\cite{WC} and the NLO polarized splitting functions (anomalous
dimensions) \cite{DGLAP} needed for the calculation of
$g_1(x,Q^2)$ in the $\rm \overline{MS}$ scheme are well known at
present. The corresponding expressions for these quantities in
the JET scheme can be found in \cite{NLO_JET}.

However, there are other contributions to $g_1$, arising from QCD
operators of higher twist (HT), namely $\tau \geq 3$, which are
related to multi-parton correlations in the nucleon. It can be
shown that these give rise to contributions to $g^N_1(x,Q^2)$ that
decrease like inverse powers of $Q^2$. The leading term has the
form $h^N(x, Q^2)/Q^2$, where $h^N(x, Q^2)$ could have a slow,
logarithmic dependence on $Q^2$.

In the kinematic regime where such terms might be relevant it is
important for consistency to realise that the QCD expression
(\ref{g1partons}) is derived under the assumption that $Q^2
>> M_N^2$. There will thus be {\it purely kinematic} corrections
to (\ref{g1partons}), which involve a power series in $M_N^2/Q^2$
with small coefficients. The leading term of these so-called
target mass corrections (TMC) therefore has a $Q^2$ behavior
similar to the genuine HT terms, but it is not a dynamical HT
effect.

In view of this we shall write
\begin{equation}
g_1(x, Q^2) = g_1(x, Q^2)_{\rm LT} + g_1(x, Q^2)_{\rm HT}~,
\label{g1QCD}
\end{equation}
where we have dropped the nucleon target label N. In (\ref{g1QCD})
\begin{equation}
g_1(x, Q^2)_{\rm LT}=g_1(x, Q^2)_{\rm pQCD} + h^{\rm TMC}(x,
Q^2)/Q^2~,
\label{LTQCD}
\end{equation}
where $h^{\rm TMC}(x, Q^2)$ is exactly calculable {\cite{TB,TMC}
and
\begin{equation}
g_1(x, Q^2)_{\rm HT}=h(x, Q^2)/Q^2~.
\label{HTQCD}
\end{equation}
As mentioned above, $h(x, Q^2)$ denotes the dynamical higher twist
power corrections to $g_1$ which represent the multi-parton
correlations in the target.
The latter are non-perturbative effects and their calculation is
model dependent (see, e.g., \cite{RenormInstanton} and references
therein). That is why a {\it model independent} extraction of the
dynamical higher twists $h(x)$ from the experimental data is
important not only for a better determination of the polarized
parton densities but also because it would lead to interesting
tests of the non-perturbative QCD regime.

\section{Method of Analysis}

In this Section we will briefly describe the method of our
analysis of the data on inclusive polarized DIS taking into
account the higher twist corrections to the nucleon structure
function $g_1^N(x, Q^2)$. In our fit to the data we have used the
following expressions for $g_1/F_1$ and $A_1$:
\begin{eqnarray}
\nonumber \left[{g_1^N(x,Q^2)\over F_1^N(x,
Q^2)}\right]_{exp}~&\Leftrightarrow&~ {{g_1^N(x,Q^2)_{\rm
LT}+h^N(x)/Q^2}\over F_2^N(x,Q^2)_{exp}}2x{[1+
R(x,Q^2)_{exp}]\over (1+\gamma^2)}~,\\
A_1^N(x,Q^2)_{exp}~&\Leftrightarrow&~{{g_1^N(x,Q^2)_{\rm LT}+
h^N(x)/Q^2}\over F_2^N(x,Q^2)_{exp}}2x[1+R(x,Q^2)_{exp}]~.
\label{g1F2Rht}
\end{eqnarray}
where $g_1^N(x,Q^2)_{\rm LT}$ is given by the leading twist
expression (\ref{LTQCD}).
In (\ref{g1F2Rht}) $h^N(x)$ are a measure of the
dynamical higher twists. In our analysis their $Q^2$ dependence
is neglected. It is small and the accuracy of the present data
does not allow to determine it. For the unpolarized structure
functions $F_2^N(x,Q^2)_{exp}$ and $R(x,Q^2)_{exp}$ we have used
the NMC parametrization \cite{NMC} and the SLAC parametrization
$\rm R_{1998}$ \cite{R1998}, respectively.

As in our previous analysis \cite{LSS2001}, for the input LO and
NLO polarized parton densities at $Q^2_0=1~GeV^2$ we have adopted
a simple parametrization
\begin{eqnarray}
\nonumber
x\Delta u_v(x,Q^2_0)&=&\eta_u A_ux^{a_u}xu_v(x,Q^2_0),\\
\nonumber
x\Delta d_v(x,Q^2_0)&=&\eta_d A_dx^{a_d}xd_v(x,Q^2_0),\\
\nonumber
x\Delta s(x,Q^2_0)&=&\eta_s A_sx^{a_s}xs(x,Q^2_0),\\
x\Delta G(x,Q^2_0)&=&\eta_g A_gx^{a_g}xG(x,Q^2_0),
\label{inputPPD}
\end{eqnarray}
where on RHS of (\ref{inputPPD}) we have used the MRST98 (central
gluon) \cite{MRST98} and MRST99 (central gluon) \cite{MRST99}
parametrizations for the LO and NLO($\rm \overline{MS}$)
unpolarized densities, respectively. The number of active flavors
is $\rm N_f=3$. The normalization factors $A_i$ in
(\ref{inputPPD}) are fixed such that $\eta_{i}$ are the first
moments of the polarized densities. To fit better the data in LO
QCD, an additional factor $(1+ \gamma_v x)$ on RHS is used for the
valence quarks. Bearing in mind that the light quark sea
densities $\Delta\bar{u}$ and $\Delta\bar{d}$ cannot be, in
principle, determined from the present inclusive data (in the
absence of polararized charge current neutrino experiments) we
have adopted the convention of a flavor symmetric sea
\begin{equation}
\Delta u_{sea}=\Delta\bar{u}=\Delta d_{sea}=\Delta\bar{d}=
\Delta s=\Delta\bar{s}.
\label{SU3sea}
\end{equation}

The first moments of the valence quark densities $\eta_u$ and $\eta_d$ are
constrained by the sum rules
\begin{equation}
a_3=g_{A}=\rm {F+D}=1.2670~\pm~0.0035~~\cite{PDG},
\label{ga}
\end{equation}
\begin{equation}
a_8=3\rm {F-D}=0.585~\pm~0.025~~\cite{AAC},
\label{3FD}
\end{equation}
where $a_3$ and $a_8$ are non-singlet combinations of the first
moments of the polarized parton densities corresponding to
$3^{\rm rd}$ and $8^{\rm th}$ components of the axial vector
Cabibbo current
\begin{equation}
a_3 = (\Delta u+\Delta\bar{u})(Q^2) - (\Delta d+\Delta\bar{d})(Q^2)~,
\label{a3ga}
\end{equation}
\begin{equation}
a_8 =  (\Delta u +\Delta\bar{u})(Q^2) + (\Delta d + \Delta\bar{d})(Q^2)
- 2(\Delta s+\Delta\bar{s})(Q^2)~.
\label{a8}
\end{equation}

The sum rule (\ref{ga}) reflects isospin SU(2) symmetry, whereas
(\ref{3FD}) is a consequence of the $SU(3)_f$ flavor symmetry
treatment of the hyperon $\beta$-decays. While the isospin
symmetry is not in doubt, there is some question about the
accuracy of assuming $SU(3)_f$ in analyzing hyperon
$\beta$-decays. We have previously studied the sensitivity of the
polarized parton densities to the deviation of $a_8$ from its
SU(3) flavor symmetric value (0.58). The results are given in
\cite{SU3}. In this analysis we will use for $a_8$ its SU(3)
symmetric value (\ref{3FD}).

In our past papers we have used the Jacobi polynomial method to
yield the structure functions $g_1^N(x,Q^2)_{\rm LT}$ from their
Mellin moments in n space. The details of this procedure are
given in \cite{LSSIJMP}. But the accuracy of this method is
limited in the low x region, $x < 0.01$, so we have now used the
inverse Mellin-transformation method (see, e.g., \cite{GRV}) which
reconstructs very precisely $g_1^N(x,Q^2)_{\rm LT}$ from its
moments in the whole $x$ region. We have repeated our fits without
including HT corrections in order to compare to our previous
results \cite{LSS2001} obtained by the Jacobi polynomial method.
We have found very good agreement between the results obtained by
both methods. The reason is that the present kinematic $x$ region
of the polarized DIS data coincides with the domain where the
Jocobi polynomial method works well. Also, the difference between
the structure functions calculated by Jacobi and inverse
Mellin-transformation methods is much smaller than the accuracy
achieved in the present polarized DIS experiments. Nevertheless,
bearing in mind its universality and, in particular, its
applicability to the semi-inclusive DIS processes, we have
decided to use the inverse Mellin-transformation approach in this
analysis.

\def\thefootnote{\dagger}
The unknown higher twists $h^N(x)$ in (\ref{g1F2Rht}) have been
extracted from the data following the method used in \cite{htF2}
and \cite{htF3} for the higher twist corrections to the
unpolarized structure functions $F_2$ and $xF_3$,
respectively\footnote{Note that the moments of the $g_1$ higher
twists have been studied in the SLAC/E143 paper \cite{g1F1data}
as well as in \cite{TMC,momHT}.} The measured $x$ region has been
split into 5 bins and to any $x$-bin two parameters $h_i^{(p)}$
and $h_i^{(n)}$ have been attached. We have found that for a
deutron the relation $h_i^{(d)}=0.925(h_i^{(p)}+ h_i^{(n)})/2$ is
a good approximation. So, to the parameters connected with the
input PPD (\ref{inputPPD}) we add the parameters $h_i^{(p)}$ and
$h_i^{(n)},~(i=1, 2, ..,5$).

All free parameters
\begin{equation}
\{~a_u,~ a_d,~ a_s,~ a_g,~ \eta_s,~\eta_g ~(\gamma_u,~\gamma_d);
~h_i^{(p)},~h_i^{(n)}~\} \label{freeparam}
\end{equation}
have been determined from the best fit to $g_1/F_1$ and $A_1$ data
using (\ref{g1F2Rht}), i.e., effectively by fitting $(g_1)_{exp}$.
Note that in the calculations of $g_1(x,Q^2)_{\rm LT}$ we have
used for the strong coupling constant $\alpha_s(Q^2)$ the same
procedure as in our previous paper \cite{LSS2001} (see the
details in LSS2001 FORTRAN code at
http://durpdg.dur.ac.uk/HEPDATA/PDF).

\section{Results}

In this section we present the numerical results of our fits to
the world data on $g_1/F_1$ \cite{g1F1data} and $A_1$
\cite{EMC,A1data}.
The data used (185 experimental points) cover the following
kinematic region:
\begin{equation}
0.005 \leq x \leq 0.75,~~~~~~1< Q^2 \leq 58~GeV^2~.
\label{kinreg}
\end{equation}

The total (statistical and systematic) errors are taken into
account. The systematic errors are added quadratically.

We prefer to discuss the results of the NLO analysis in the JET
(or so-called chirally invariant) factorization scheme
\cite{JET}. In this scheme the first moment of singlet $\Delta
\Sigma(Q^2)$, as well as the strange sea polarization $(\Delta
s+\Delta\bar{s})(Q^2)$, are ${\it Q^2}$ {\it independent}
quantities. Then, it is meaningful to directly interpret $\Delta
\Sigma$ as the contribution of the quark spins to the nucleon
spin and to compare its value obtained from the DIS region with
the predictions of the different (constituent, chiral, etc.) quark
models at low $Q^2$. Later we will briefly comment on the scheme
dependence effects on the results of the analysis.

\subsection{Higher twist effects}

The numerical results of our fits to the data are summarized in
Tables 1 and 2. As seen from the values of $\chi^2$ per degree of
freedom $\chi^2_{\rm DF}$ in Table 1, a very good description of
the data is achieved. The best LO and NLO(JET) fits correspond to
$\chi^2_{\rm DF,LO}=0.892$ and to $\chi^2_{\rm DF,NLO}=0.858$. We
have found that the fit to the data is significantly improved,
especially in the LO case, when the higher twist corrections to
$g_1^N$ are included in the analysis (see Table 2). In contrast
to the case when the HT corrections to $g_1$ are not taken into
account in the fits, the value of $\chi^2_{\rm LO}(\rm HT)$ is
very close to that of $\chi^2_{\rm NLO}(\rm HT)$, which is an
indication that the tail of the neglected higher order
logarithmic corrections to $g_1$ resemble a power behavior of
order ${\cal O}(1/Q^2)$ \cite{Forte}. A similar behavior of
$\chi^2$ has been observed in the QCD analysis of the unpolarized
structure function $xF_3(x,Q^2)$ in \cite{htF3}.

The extracted higher twist corrections to the proton and neutron
spin structure functions, $h^p(x)$ and $h^n(x)$, are shown in
Fig. 2. As seen from Fig. 2 the corrections for the proton and
neutron have a different shape. While $h^p(x)$ changes sign in
the LO as well in the NLO case, $h^n(x)$ is non-negative in the
measured $x$ region in both cases. One can see also that the HT
corrections to the proton structure function $g_1^p$ appear to be
smaller when for $(g_1)_{\rm LT}$ the NLO approximation is used.
In Fig. 3 we demonstrate how the choice of the factorization
scheme for the perturbative calculation of $(g_1)_{\rm LT}$
influences the higher twists results. The results are presented
for the JET and $\rm \overline{MS}$ schemes. It is seen that the
HT corrections to $g_1$ in both cases coincide within the errors.
\def\thefootnote{\dagger}
The small deference between the central values could be considered
as an estimation of the NNLO effects in $(g_1)_{\rm LT}$\footnote{
In \cite{BB} the HT terms have been discussed using for them two
phenomenological parametrizations. The authors conclude that they
do not find a significant higher twist contribution to $g_1$ in
an NLO treatment of $(g_1)_{\rm LT}$. On other hand, studying the
higher twist effects in the moments of $g_1$, it was shown in
\cite{TMC} that while the first moment of higher twist is quite
small, the higher order moments are relevant at $Q^2 \sim~$few
$GeV^2$.}.

\subsection{NLO polarized parton densities}

Let us discuss now the polarized parton densities extracted from
the data in the presence of the HT corrections to $g_1$. We will
call this set of parton densities PD($g_1^{\rm LT}+\rm HT$). In
Fig. 4 we compare the NLO(JET) polarized PD($g_1^{\rm NLO}+\rm
HT$) with those obtained in our analysis \cite{LSS2001} where we
performed fits to the data according to (\ref{g1F1method}). We
will call the latter PD($g_1^{\rm NLO}/F_1^{\rm NLO}$). As seen
from Fig. 4 the two sets of polarized parton densities are very
close to each other. This is a good illustration of the fact that
a fit to the data on $A_1(\sim g_1/F_1~)$ using for the $g_1$ and
$F_1$ structure functions their NLO leading twist expressions
($\chi^2_{\rm DF,NLO}=0.859$) is equivalent to a fit to the $g_1$
data taking into account the higher twist corrections to $g_1$
($\chi^2_{\rm DF,NLO}=0.858$). In other words, this analysis
confirms once more that the higher twist corrections to $g_1$ and
$F_1$ approximately cancel in the ratio $g_1/F_1$.

\subsection{LO polarized parton densities}

Let us turn now to the LO polarized parton densities. In LO QCD
$\Delta G(x,Q^2)$ does not contribute directly to $g_1$ and the
gluons cannot be determined from DIS data alone. For this reason
the LO fit to the data was performed using for the input
polarized gluon density $\Delta G(x,Q^2_0)$ the one extracted in
the NLO fit to the data:
\begin{equation}
\Delta G(x,Q^2_0)_{\rm LO}= \Delta G(x,Q^2_0)_{\rm NLO(JET)}~.
\label{LOdelG_NLO}
\end{equation}
It is important to note that in the polarized case the LO
approximation has some peculiarities compared to the unpolarized
one. As a consequence of the gluon axial anomaly, the difference
between NLO anti-quark polarizations $\Delta\bar {q}_i$ in
different factorization schemes can be quite large, comparable in
magnitude to the $\Delta\bar {q}_i$ themselves (see, e.g.,
\cite{LSS2001}). In this case the leading order will be a bad
approximation, at least for the polarized sea-quark densities.
Also, bearing in mind that in polarized DIS most of the data
points are at low $Q^2$, lower than the usual cuts in the
analyses of unpolarized data ($Q^2\geq 4-5~GeV^2$), the NLO
corrections to all polarized parton densities are large in this
region and it is better to take them into account. Nevertheless,
the LO polarized parton densities may be useful for some practical
purposes; e.g., for preliminary estimations of the cross sections
in future polarized experiments, etc. They are also needed for
comparison with those extracted from semi-inclusive DIS data
\cite{SIDIS}, where the NLO QCD analysis is still very
complicated. The extracted LO polarized parton densities
PD($g_1^{\rm LO}+HT$) are shown in Fig. 5. Also shown in Fig. 5
are the LO polarized PD($g_1^{\rm LO}/F_1^{\rm LO}$) obtained in
our analysis \cite{LSS2001}. In contrast to the NLO case, the two
sets of LO polarized densities are significantly different. As a
result we obtain different theoretical curves for $g_1$ (see Fig.
6). To illustrate how these curves fit the data, the SLAC/E143
experimental proton data at $Q^2=5~GeV^2$ are also shown. As seen
from Fig. 6, the proton structure function ${g_1^p(x, Q^2)}_{LO}$
calculated using the LO polarized PD($g_1^{\rm LO}/F_1^{\rm LO}$)
does not agree with the data for $x < 0.25$. Note that at the
same time the ratio $g_1^{\rm LO}/F_1^{\rm LO}$ fits the world
data on $g_1/F_1$ and $A_1$ quite well in the measured $(x, Q^2)$
region ($\chi^2_{\rm DF}=0.921$) \cite{LSS2001}. The main reason
for this peculiarity is that the LO approximations for the
unpolarized structure functions $F_2(x, Q^2)$ and $F_1(x, Q^2)$
(or $R(x, Q^2))$ presented in the literature are not self
consistent. The unpolarized parton densities in leading order QCD
(more correctly in leading logarithmic approximation LLA) are
usually extracted including in the data set of the analysis only
the experimental data on $F_2$ and ignoring the data on R (or
F1). Remember that in LLA of QCD the structure functions satisfy
the Callan-Gross relation $2xF_1(x,Q^2)=F_2(x, Q^2)$ which leads
to $R=4M^2x/Q^2$. However, the experimental data on $F_2$ and $R$
do not satisfy these relations in a large kinematic region. They
are approximately satisfied only at large $x$ and/or large $Q^2$.
At small $x$ and $Q^2$ the experimental values of $F_2$ are larger
than those of $2xF_1$ by up to 30\%. That is why the extracted
sets of LO unpolarized parton densities (MRST, CTEQ, etc.) are
not quite consistent. While they fit well the data on $F_2$, they
badly fail to describe the $R(F_1)$ data in the region of small
$x$ and $Q^2$. One way to improve the situation would be to
perform a LO QCD fit including in the data set the $R(F_1)$ data
too. Also, if the data at low $Q^2$, lower than the usual cuts
($Q^2\geq 4-5~GeV^2$), are included in the analysis, the higher
twist corrections to $F_2$ and $F_1$ should be taken into account.

\section{Conclusion}

We have analyzed the world data on inclusive polarized deep
inelastic lepton-nucleon scattering in leading and
next-to-leading order of QCD including in the analysis the higher
twist $h^N(x)/Q^2$ and the target mass corrections to the nucleon
spin structure function $g_1^N(x, Q^2)$. We have found that the
fit to the data on $g_1$ is essentially improved, especially in
the LO case, when the higher twist terms are included in the
analysis. The $x$-dependence of the higher twists $h^N(x)$ have
been extracted from the data in a model independent way. It is
shown that the size of their contribution to $g_1$ is not
negligible and their shape depends on the target:$~h^p(x)$
changes sign while $h^n(x)$ is a non-negative function in the
measured $x$ region.

We have found that the polarized parton densities depend on
whether the higher twist terms are or are not included in the
analysis of $g_1$. Moreover, the NLO polarized parton densities
extracted from the $g_1$ data in the presence of higher twist
terms are in good agreement with those determined by our previous
fits \cite{LSS2001} to the data on $g_1/F_1$ and $A_1$ using for
the structure functions $g_1$ and $F_1$ only their {\it leading}
twist expressions in NLO QCD. This observation confirms once more
that the higher twist corrections to $g_1/F_1$ and $A_1$ are
negligible so that in the analysis of $g_1/F_1$ and $A_1$ data it
is enough to account only for the leading twist of the structure
functions $g_1$ and $F_1$. On the other hand, in fits to the $g_1$
data themselves the higher twist contribution to $g_1$ must be
taken into account. The latter is especially important for the
LO QCD analysis of the inclusive and semi-inclusive DIS data. \\

\vskip 4mm
{\large \bf Acknowledgments}
\vskip 4mm

This research was supported by a UK Royal Society Collaborative
Grant, by the JINR-Bulgaria Collaborative Grant, by the RFBR (No
00-02-16696), INTAS 2000 (No 587) and by the Bulgarian National
Science Foundation under Contract Ph-1010.\\


\newpage

\vskip 0.6 cm
\begin{center}
\begin{tabular}{cl}
&{\bf Table 1.} Parameters of the LO and NLO(JET) input parton
densities at \\
&$Q^2=1~GeV^2$ as obtained from the best fits to the world
$g_1/F_1$ and $A_1^N$ data \\
&including the HT corrections to $g_1$. The errors shown are total
(statistical and \\
&systematic). The parameters marked by (*) are fixed. Note that
the TMC are\\
&included in $(g_1)_{\rm LT}$.
\end{tabular}
\vskip 0.6 cm
\begin{tabular}{|c|c|c|c|c|c|c|} \hline
    Fit &~$(g_1)_{LO}+h(x)/Q^2$~&~~$(g_1)_{NLO}+h(x)/Q^2$~~\\ \hline
 $\rm DF$        &  185~-~16          &     185~-~16 \\
 $\chi^2$        &  150.7             &      145.0   \\
 $\chi^2/\rm DF$ &  0.892             &      0.858   \\  \hline
 $\eta_u$        &~~0.926$^*$         &    $0.926^*$    \\
 $a_u$           &~~0.000~$\pm$~0.002~&~~0.312~$\pm$~0.048~\\
 $\gamma_u$      &~~1.556~$\pm$~0.261~&    $0^*$          \\
 $\eta_d$        &-~0.341$^*$         &    $-0.341^*$      \\
 $a_d$           &~~0.000~$\pm$~0.005~&~~0.000~$\pm$~0.049~  \\
 $\gamma_d$      &~~2.808~$\pm$~1.249~&    $0^*$            \\
 $\eta_s$        &-~0.072~$\pm$~0.008~&-~0.045~$\pm$~0.007~ \\
 $a_s$           &~~0.601~$\pm$~0.064~&~~1.583~$\pm$~0.434~ \\
 $\eta_g$        &~~$0.803^*$~        &~~0.803~$\pm$~0.244~ \\
 $a_g$           &~~$0.376^*$~        &~~0.376~$\pm$~0.503  \\ \hline
 $x_i$           & \multicolumn{2}{|c|}{$h^p(x_i)~[GeV^2]$}   \\  \hline
  0.028          &~~0.013~$\pm$~0.036    &~~0.064~$\pm$~0.044   \\
  0.100          &-~0.076~$\pm$~0.032    &-~0.007~$\pm$~0.034 \\
  0.200          &-~0.145~$\pm$~0.032    &-~0.060~$\pm$~0.035 \\
  0.350          &-~0.030~$\pm$~0.035    &-~0.008~$\pm$~0.038 \\
  0.600          &~~0.035~$\pm$~0.019    &~~0.026~$\pm$~0.021 \\ \hline
 $x_i$           &    \multicolumn{2}{|c|}{$h^n(x_i)~[GeV^2]$} \\ \hline
  0.028          &~~0.234~$\pm$~0.073    &~~0.178~$\pm$~0.078 \\
  0.100          &~~0.192~$\pm$~0.048    &~~0.199~$\pm$~0.050 \\
  0.200          &~~0.035~$\pm$~0.056    &~~0.079~$\pm$~0.059 \\
  0.325          &~~0.072~$\pm$~0.071    &~~0.055~$\pm$~0.073 \\
  0.500          &~~0.023~$\pm$~0.043    &-~0.020~$\pm$~0.040 \\ \hline
\end{tabular}
\end{center}
\vskip 2.0 cm

\vskip 0.6 cm
\begin{center}
\begin{tabular}{cl}
&{\bf Table 2. }The values of $\chi^2$ for the LO and NLO QCD
fits without HT included \\
&compared to those when the HT corrections to $g_1$ are taken
into account. The \\
&TMC are included in $(g_1)_{\rm LT}$ .
\end{tabular}
\vskip 0.6 cm
\begin{tabular}{|c|c|c|c|c|c|c|} \hline
    Fit &~~LO(HT=0)~~&~~NLO(HT=0)~~&~~LO+HT~~&~~NLO+HT~~\\ \hline
 $\chi^2$        &  244.5     &  218.8   &  150.9   &  145.0 \\
 $\rm DF$        &  185~-~6  &  185~-~6 & 185~-~16 & 185~-~16  \\
 $\chi^2/\rm DF$ &  1.36      &  1.22    &  0.893   &  0.858  \\   \hline
\end{tabular}
\end{center}
\vskip 2.0 cm

\newpage
\noindent {{\bf Figure Captions}} \vskip 5mm \noindent {\bf
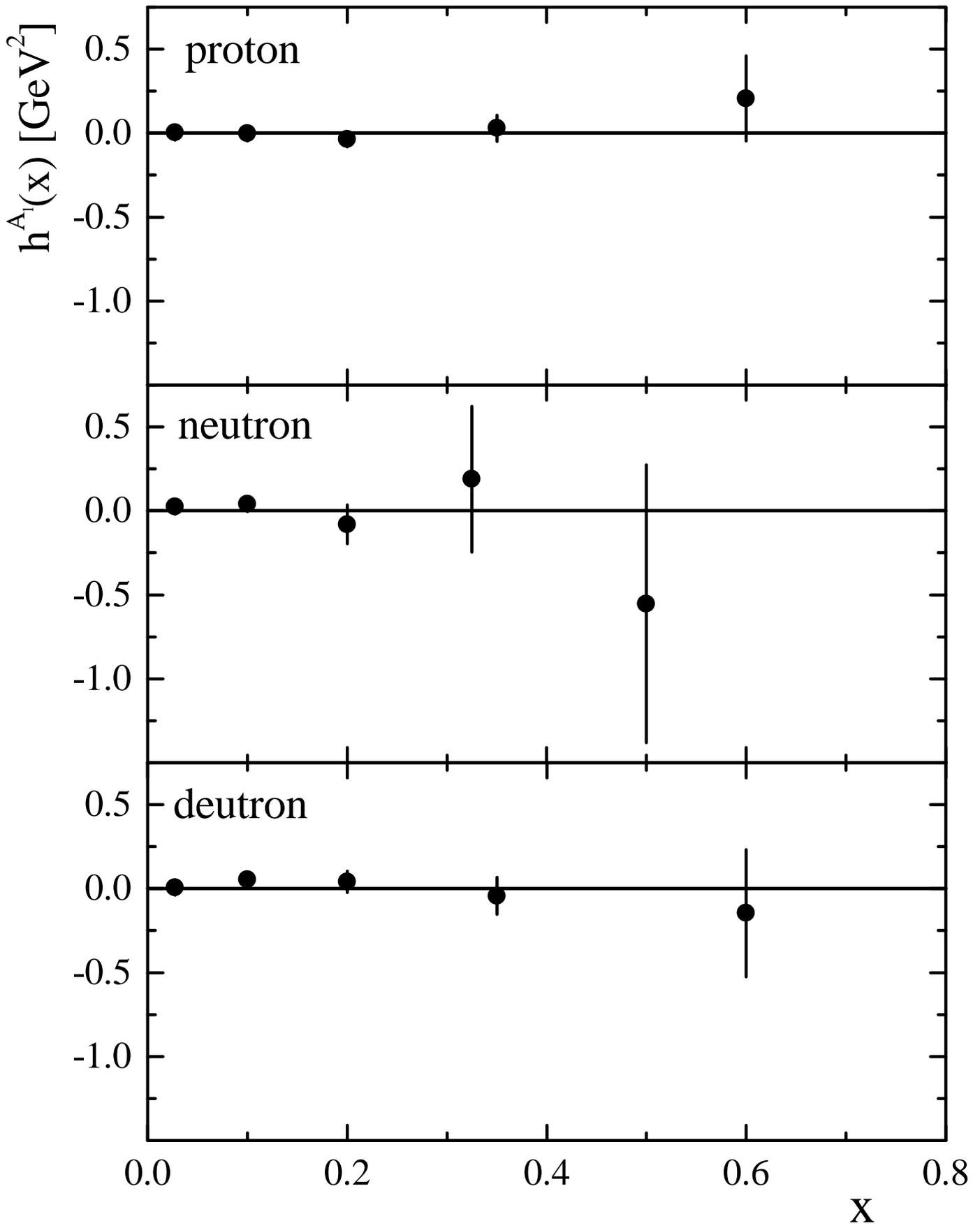
Fig.1.} Effective higher twist contribution $h^{A_1}(x)$ to the
spin asymmetry $A_1^N(x,Q^2)$ extracted from the data. Compared
to our 1999 result (Fig. 4 \cite{LomConf} and Fig. 1
\cite{LSS2001}) we present here the results of a new analysis
including in the data
set the SLAC/155 proton data not available at that time. \\

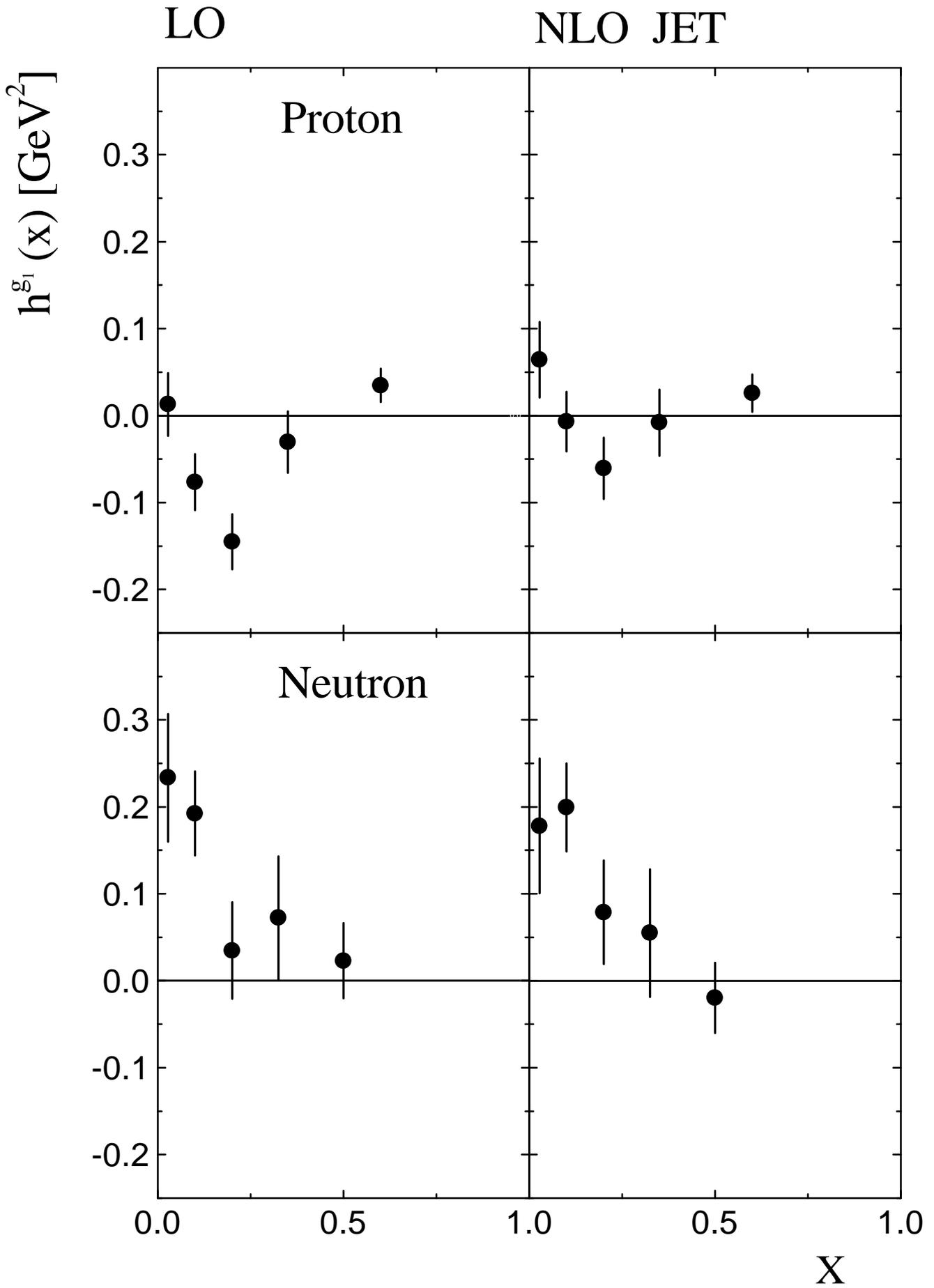
\noindent {\bf Fig. 2.} Higher twist corrections to the proton
and neutron $g_1$ structure functions extracted from the data on
$g_1$ in the case of LO and NLO QCD approximation for
$g_1(x,Q^2)_{\rm LT}$.\\

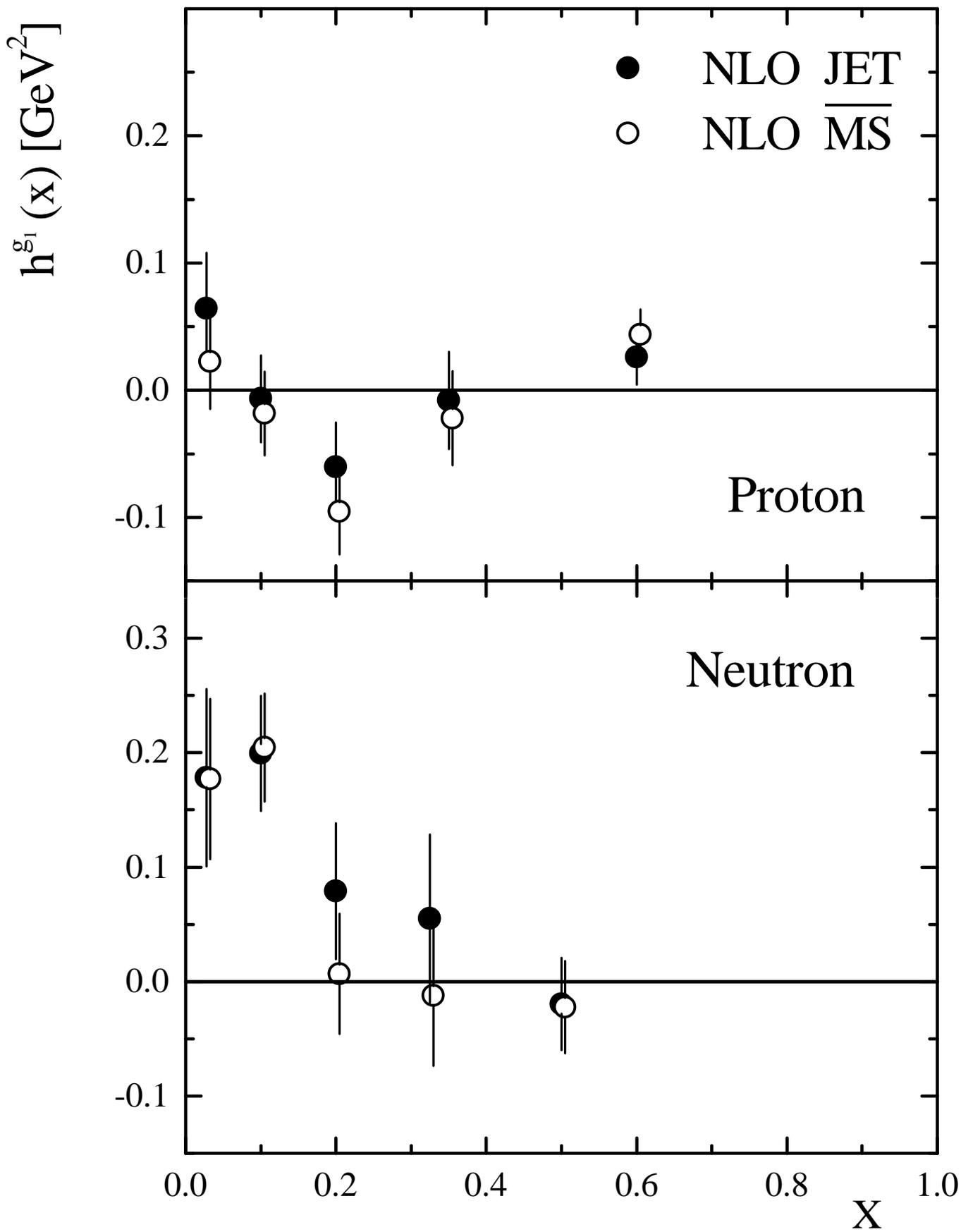
\noindent {\bf Fig. 3.} Higher twist corrections to the proton
and neutron $g_1$ structure functions extracted from the data
when the leading twist calculations of $g_1(x,Q^2)_{\rm NLO}$ are
performed in different factorization schemes.\\

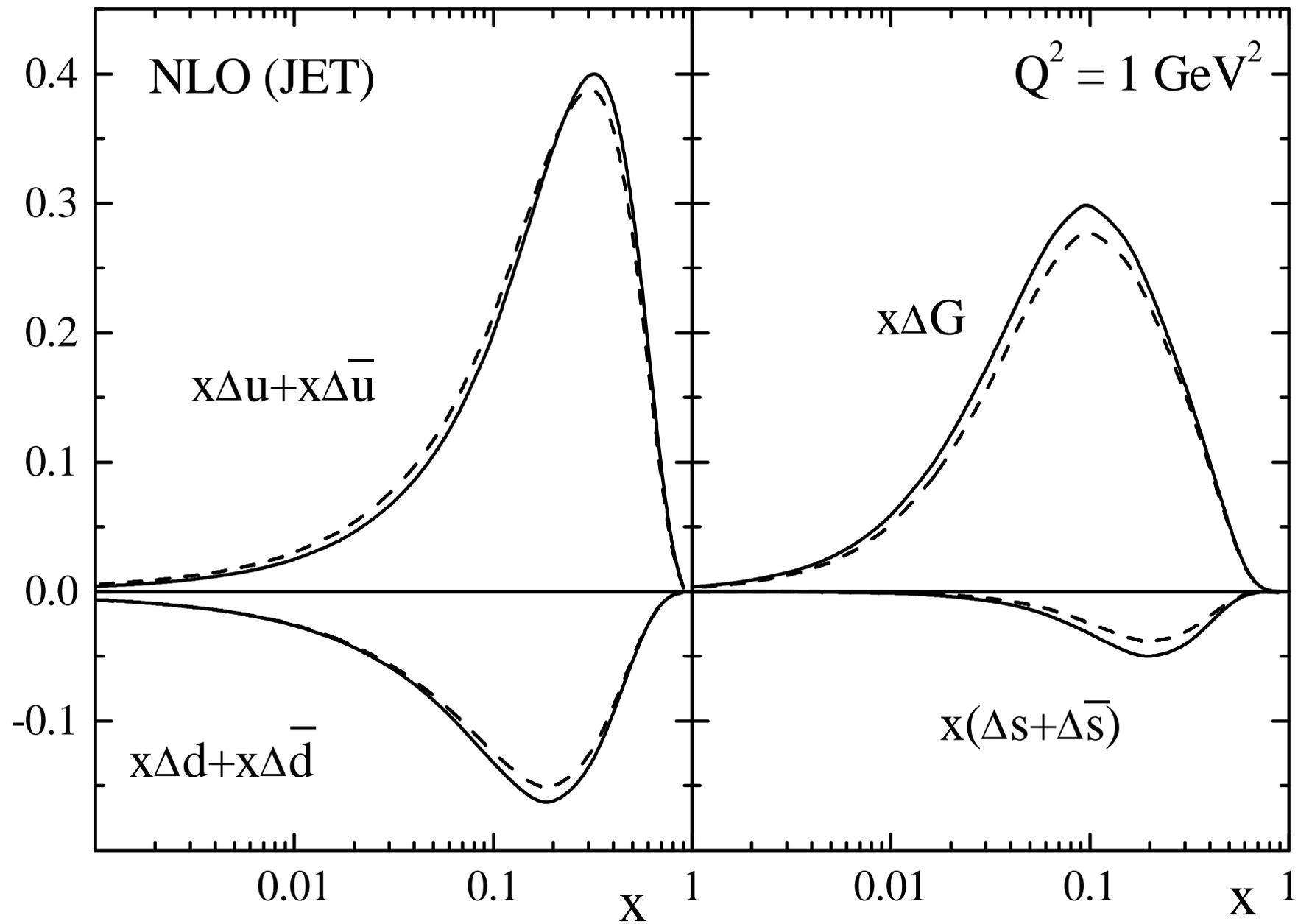
\noindent {\bf Fig. 4.} NLO(JET) polarized parton densities
PD($g_1^{\rm NLO}+\rm HT$) (solid curves) compared to PD($g_1^{\rm
NLO}/F_1^{\rm NLO}$) (dashed curves) at $Q^2=1~GeV^2$ (see the text).\\

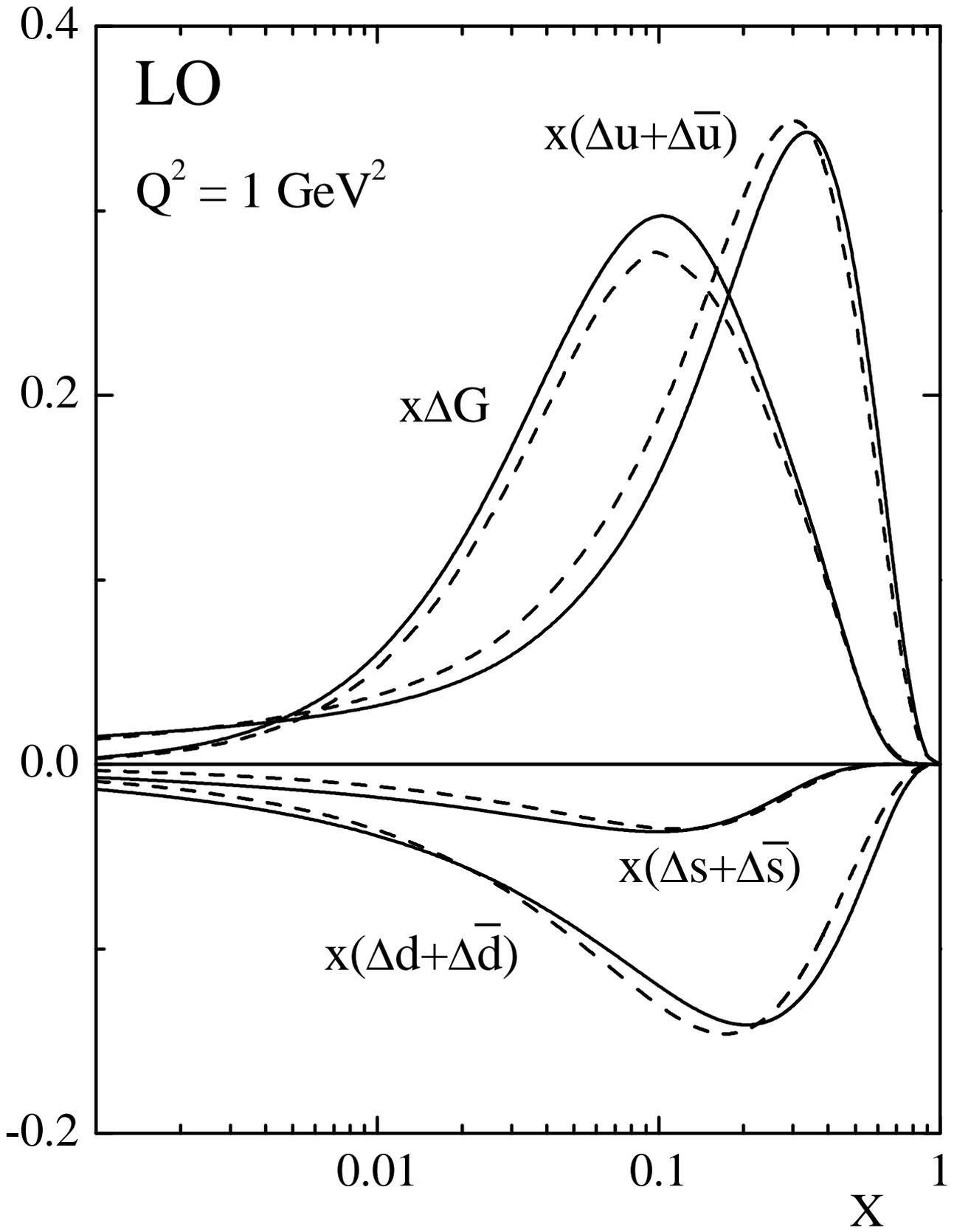
\noindent {\bf Fig. 5.} LO polarized parton densities PD($g_1^{\rm
LO}+\rm HT$) (solid curves) compared to PD($g_1^{\rm
LO}/F_1^{\rm LO}$) (dashed curves) at $Q^2=1~GeV^2$ (see the text).\\

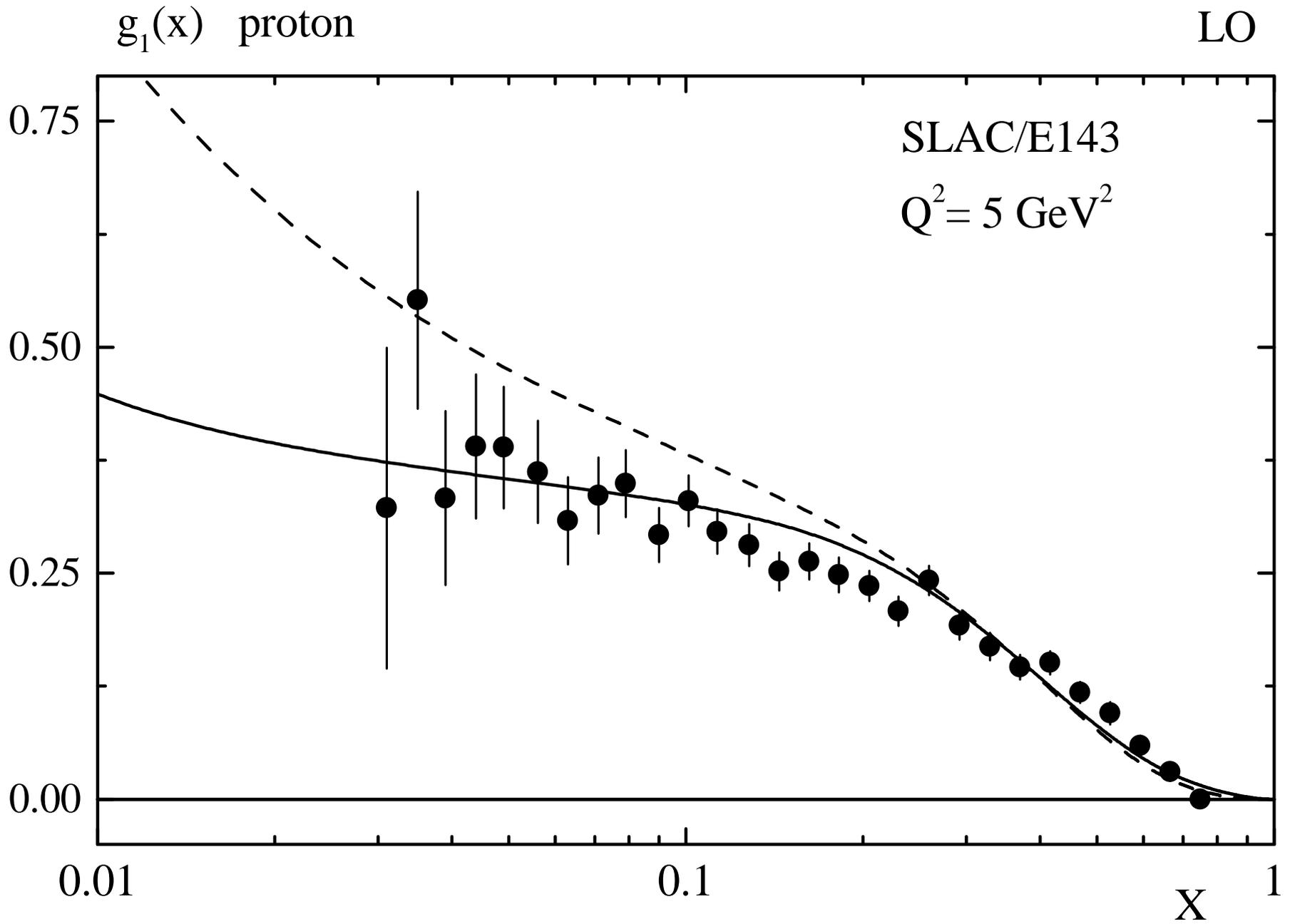
\noindent {\bf Fig. 6.} Comparison of the proton structure
function $g_1^{\rm LO}$ calculated using the polarized parton
densities PD($g_1^{\rm LO}+\rm HT$) (solid curve) and PD($g_1^{\rm
LO}/F_1^{\rm LO}$) (dashed curve) with SLAC/E143 proton data.

\end{document}